\documentclass[reprint,
 amsmath,amssymb,
 aps,
]{revtex4-2}
\usepackage{booktabs}
\usepackage{xcolor}
\usepackage{graphicx}% Include figure files
\usepackage{dcolumn}% Align table columns on decimal point
\usepackage{bm}% bold math
\usepackage{hyperref}% add hypertext capabilities
\usepackage{babel}
\usepackage{comment}
\usepackage{soul}
\usepackage{tikz-feynman}
\usepackage{tikz}
\usepackage[mathscr]{euscript}
\usepackage{etoolbox}
\apptocmd{\sloppy}{\hbadness 10000\relax}{}{}
\newcommand{\zbar}{\bar{z}}

\newcommand{\Cee}{\mathscr{C}}

\begin{document}
\preprint{APS/123-QED}
\title{Conformal Invariance and Multifractality at Anderson Transitions in Arbitrary Dimensions}

\author{Jaychandran Padayasi}
\author{Ilya Gruzberg}%
% \email{gruzberg.1@osu.edu}
\affiliation{%
Department of Physics, Ohio State University, 191 West Woodruff Avenue, Columbus, Ohio, 43210, USA
}%

\date{\today}% It is always \today, today,
             %  but any date may be explicitly specified

\begin{abstract}

Multifractals arise in various systems across nature whose scaling behavior is characterized by a continuous spectrum of multifractal exponents $\Delta_q$. In the context of Anderson transitions, the multifractality of  critical wave functions is described by operators $O_q$ with scaling dimensions $\Delta_q$ in a field-theory description of the transitions. The operators $O_q$ satisfy the so-called Abelian fusion expressed as a simple operator product expansion. Assuming conformal invariance and Abelian fusion, we use the conformal bootstrap framework to derive a constraint that implies that the multifractal spectrum $\Delta_q$ (and its generalized form) must be quadratic in its arguments in any dimension $d \geq 2$.

\end{abstract}

\maketitle

Multifractal (MF) measures with intricate scaling arise in such diverse subjects as dy\-na\-mi\-cal chaos~\cite{Halsey-Fractal-1986, Paladin-Anomalous-1987}, weather and climate~\cite{Lovejoy-The-weather-2013}, turbulence~\cite{Mandelbrot-Intermittent-1974, Frisch-Fully-1980, Frisch-Turbulence-1985, Meneveau-The-multifractal-1991, Benzi-Multifractal-2022}, fractal growth~\cite{Vicsek-Fractal-1992, Bauer-2D-growth-2006, Meakin-Fractals-2011, Bunde-Fractals-2012}, critical clusters in statistical mechanics~\cite{Cates-Diffusion-1987, Gruzberg-Stochastic-2006, Duplantier-Conformal-2006}, disordered magnets and other random critical points~\cite{Ludwig-Infinite-1990, Monthus-Symmetry-2009}, Anderson transitions (ATs)~\cite{Evers2007, Hof-Calculation-1986, Wegner-Anomalous-I-1987, Wegner-Anomalous-II-1987, Burmistrov-Multifractality-2013, Burmistrov-Tunneling-2014, Burmistrov-Multifractality-2015, Repin-Mesoscopic-2016}, mathematical finance~\cite{Bouchaud-Apparent-2000, Bacry-Continuous-2008}, random energy landscapes~\cite{Fyodorov-Multifractality-2010, Fyodorov-High-values-2015}, Gaussian multiplicative chaos~\cite{Rhodes-Gaussian-2013}, and rigorous approaches to conformal field theory (CFT)~\cite{Rhodes-Lecture-2016, Vargas-Lecture-2017}. 

A MF measure $\mu(\mathbf{r})$ is characterized by the scaling of its moments with the system size $L$: $\int d^dr\ \mu^q(\mathbf{r}) \sim L^{-\tau_q}$, with a continuum of exponents $\tau_q$ that depend \emph{nonlinearly} on $q$. MF moments $\mu^q(\mathbf{r})$ can be represented by local operators $O_q(\mathbf{r})$ in a scale-invariant field theory, with scaling dimensions, also called the \textit{MF spectrum}, $\Delta_q \equiv \tau_q - d(q-1) +  q\Delta_1$~\cite{Duplantier-Multifractals-1991}.  

Similar to critical phenomena, one may expect the scale invariance to be enhanced to conformal invariance (though this is not guaranteed~\cite{Riva-Cardy-2005, Nakayama-review-2015}), in which case, MF properties can be described by a CFT. Our main result is that in this situation, and under the assumption of \emph{Abelian fusion} [see Eq.~\eqref{Abelian-fusion}] that is valid for ATs, in \emph{any dimensionality} $d \geq 2$, the MF spectrum $\Delta_q$ \emph{must be parabolic}, see Eq.~\eqref{parabolic-spectrum} below. Our result is general and should apply to all MF measures that obey conformal invariance and Abelian fusion.

Our work is motivated by and of particular significance to the study of MF wave functions at ATs\cite{Evers2007}~\footnote{In the literature on ATs, the MF spectrum that we denote here as $\Delta_q$ is usually denoted by $x_q$.}, where the parabolicity of $\Delta_q$ was predicted in a $d = 2$ CFT~\cite{Bondesan2016}. This prediction was tested analytically and numerically, and was found to be violated at two-dimensional (2D) ATs in various symmetry classes~\cite{Obuse-Boundary-2008, Evers-Multifractality-2008, Puschmann-Quartic-2021, Karcher2021, Karcher2022, Karcher2022b, Karcher-Generalized-2023}. This has led to the un\-der\-stan\-ding that conformal invariance might be lost at these critical points. Similarly, numerical studies of multifractality in $d = 3,4,5$ have found strong deviations from parabolicity~\cite{Rodriguez-Multifractal-2011, Ujfalusi-Finite-size-2015, Lindinger-Multifractal-2017, Tarquini-Critical-2017} but there has not been any prediction in $d > 2$ from a CFT perspective. Our Letter provides such a prediction. 

{\it Multifractals and field theory.}---We first recall properties of MF spectra that follow from general principles. The function $\tau_q$ is nondecreasing and convex, which implies the existence of $q_* > 0$ such that $\Delta_{q_*} = 0$~\cite{Gruzberg-Symmetries-2011}.  Further constraints follow from studying \textit{MF correlators}  in a field theory via the relation~\cite{Duplantier-Multifractals-1991, Eyink-Multifractals-1995}
\begin{align}
    \overline{\mu^{q_1}(\mathbf{r}_1) \ldots \mu^{q_n}(\mathbf{r}_n)}
    \propto
   \big\langle O_{q_1}(\mathbf{r}_1) \ldots O_{q_n}(\mathbf{r}_n) \big\rangle.
\end{align}
The overbar denotes spatial or disorder average, while the angular brackets denote a field-theory expectation value.  

Of crucial importance is the additive, or \emph{Abelian}, nature of the operator product expansion (OPE) of  two MF operators $O_{q_1}$ and $O_{q_2}$~\cite{Cates-Spatial-1987, Duplantier-Multifractals-1991, Eyink-Multifractals-1995}:
\begin{align}
    O_{q_1}(\mathbf{r}) O_{q_2}(0) \propto |\mathbf{r}|^{\Delta_{q_1 + q_2} - \Delta_{q_1} - \Delta_{q_2}} O_{q_1 + q_2}(0) + \ldots, 
    \label{Abelian-OPE}
\end{align}
where the ellipsis denotes subleading operators. As a consequence, a MF correlator $\langle \prod_i O_{q_i}(\mathbf{r}_i) \rangle$  scales as $L^{-\Delta_{q_1 + q_2 + \ldots}}$ in the infrared. In the $L \rightarrow \infty$ limit,  only \textit{charge-neutral correlators} with $\Delta_{q_1 + q_2 + \ldots} = 0$ can be studied by field-theory methods~\cite{Karcher2021}.  

When conformal invariance is present, it fixes two-point functions: $\langle O_{q_1}(\mathbf{r})O_{q_2}(0)\rangle = \delta_{\Delta_{q_1},\Delta_{q_2}}|\mathbf{r}|^{-2\Delta_{q_1}}$.  This form is consistent with the OPE~\eqref{Abelian-OPE} if $\Delta_{q_1 + q_2} = 0$ and $\Delta_{q_1} = \Delta_{q_2}$.  Given the convexity of the MF spectrum, $\Delta_q \ne \Delta_{-q}$, and the only consistent choice is $q_2 = q_* - q$. Then we get the symmetry relation~\cite{Gruzberg-Symmetries-2011}
\begin{equation}
\label{eq:SymmetryRelation}
    \Delta_q = \Delta_{q_* - q}.
\end{equation}
More generally, only MF corelators with $\sum_i q_i = q_*$ are consistent with conformal invariance~\cite{Karcher2021}. The relation~\eqref{eq:SymmetryRelation} is on more rigorous footing for ATs, where it follows from the Weyl symmetry~\eqref{Weyl-symmetry} of the critical theory and does not rely on conformal invariance.

\textit{Multifractality at ATs.}---ATs between metals and insulators, as well as between topologically distinct localized phases, are a major focal point in the study of disordered systems~\cite{Evers2007}. Critical properties at ATs are notoriously difficult to study because of the strongly coupled nature of the critical points. 

A remarkable property of ATs is the multifractality of critical wave functions, or the local density of states $\nu(\mathbf{r})$ whose moments scale as $\overline{\nu^{q}(\mathbf{r})} \sim L^{-\Delta_q}$. There are more general combinations $P_\gamma$ of critical wave functions~\cite{Gruzberg-Classification-2013, Karcher2021, Karcher2022, Karcher2022b} labeled by vectors $\gamma = (q_1,\ldots,q_n)$ of complex numbers $q_i$, with scaling dimensions $\Delta_\gamma$.  MF properties at ATs are under better control than in general multifractals, since they can be rigorously established within the field theories of ATs, the nonlinear sigma models on cosets $\mathcal{G}/\mathcal{K}$ of certain Lie supergroups~\cite{efetov1983supersymmetry, efetov1997supersymmetry, mirlin00, Evers2007, wegner2016supermathematics}. In these models, $P_\gamma$ are represented by gradientless composite operators $O_\gamma$~\cite{Hof-Calculation-1986, Wegner-Anomalous-I-1987, Wegner-Anomalous-II-1987, Gruzberg-Classification-2013}. A key fact is that $O_\gamma$ can be constructed as highest-weight vectors under the action of the Lie superalgebra of $\mathcal{G}$ with weights $\gamma$~\cite{Gruzberg-Classification-2013}. Then, the $\mathcal{G}$ symmetry of the target space (assumed not broken at the critical point) implies Abelian fusion
\begin{align}
    O_{\gamma_1} \times O_{\gamma_2} \sim O_{\gamma_1 + \gamma_2}
    + \ldots,
    \label{Abelian-fusion}
\end{align}  
where the ellipsis denotes now derivatives of $O_{\gamma_1 + \gamma_2}$ and not general subleading operators as in Eq.~\eqref{Abelian-OPE}.

The $\mathcal{G}$ symmetry also leads to the \emph{Weyl symmetry} of the MF spectra $\Delta_\gamma = \Delta_{w\gamma}$,  $w \in W$~\cite{Gruzberg-Classification-2013}. The Weyl group $W$ acts in the space of weights $\gamma$ and is generated by 
\begin{align}
   q_i &\rightarrow -c_i - q_i, 
   &
   q_i &\rightarrow q_j+(c_j-c_i)/2.
  \label{Weyl-symmetry}
\end{align}

The coefficients $c_i$ of the half-sum of the positive roots $\rho_b = \sum_{j=1}^{n} c_j e_j$ in a standard basis $e_j$ are known for all families of symmetric superspaces~\cite{Mirlin-Exact-2006, Gruzberg-Symmetries-2011, Gruzberg-Classification-2013, Karcher2021, Karcher2022, Karcher2022b, Karcher-Generalized-2023}. The Weyl symmetry implies the existence of the operator $O_{-\rho_b}$ with vanishing scaling dimension $\Delta_{-\rho_b} = 0$. The corresponding neutrality condition for generalized MF correlators $\langle \prod_i O_{\gamma_i} \rangle$ is $\sum_i \gamma_i = -\rho_b$. The simple MF operators $O_q$ and the spectrum $\Delta_q$ corresponds to $\gamma = (q, 0, 0, \ldots, 0)$. In this case  $c_1 = - q_*$, and the Weyl symmetry reduces to Eq.~\eqref{eq:SymmetryRelation}. 
%As we mentioned, in the context of ATs, the relation~\eqref{eq:SymmetryRelation} does not require conformal invariance, in contrast to generic MF spectra. 

The Weyl symmetry is fully supported by numerical and analytical results for various symmetry classes and dimensions $d \ge 2$~\cite{Evers-Multifractality-2003, Mirlin-Wavefunction-2003, Mirlin-Exact-2006, Mildenberger-Boundary-2007, Mildenberger-Wave-2007, Obuse-Multifractality-2007, Obuse-Boundary-2008, Evers-Multifractality-2008, Vasquez-Multifractal-2008, Rodriguez-Multifractal-2008, Rodriguez-Multifractal-2011, Karcher2021, Karcher2022, Karcher2022b, Karcher-Generalized-2023}.

\textit{Multifractality and CFT in $d=2$.}---2D CFTs possess the infinite-dimensional Virasoro symmetry. In this setting, the ellipsis in Eq.~\eqref{Abelian-fusion} represents Virasoro descendants, and leads to a single Virasoro block in a four-point function of MF operators, and a Vafa-Lewellen~\cite{Vafa1988, Lewellen1988} constraint on the MF spectra. The unique solution of this constraint subject to the symmetries~\eqref{Weyl-symmetry} is the parabolic spectrum~\cite{Bondesan2016, Zirnbauer-The-integer-2019, Karcher2021}
\begin{align}
    \Delta_\gamma &= - b \textstyle\sum_i q_i (q_i + c_i),
&
    \Delta_q &= b q (q_* - q),
    \label{parabolic-spectrum}
\end{align}
where the parameter $b$ cannot be determined from symmetry considerations alone. The second equation above is the simplification of the first for the simple MF spectrum $\Delta_q$. In $d=2$, the MF operators appear as vertex operators in a \textit{Coulomb gas theory} (a Gaussian free field with a background charge)~\cite{Bondesan2016}. 

The central result of this Letter is that, once we assume conformal invariance and Abelian fusion, Eqs.~\eqref{parabolic-spectrum} hold for MF spectra at ATs in any dimension $d \geq 2$.

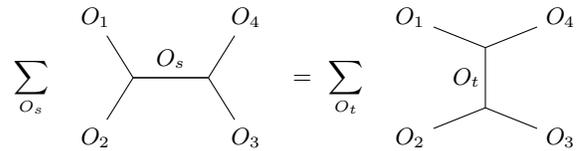
\begin{figure}
$\displaystyle \sum_{O_s}$
\begin{minipage}{0.35\columnwidth}
\begin{tikzpicture}
  \begin{feynman}
    % Incoming operators
    \vertex (a) at (-1.0, 0.8) 
    %{$O_{q_1}$};
    {$O_1$};
    \vertex (b) at (-1.0, -0.8) 
    %{$O_{q_2}$};
    {$O_2$};
    % Intermediate vertices
    \vertex (c) at (-0.5, 0) ;
    \vertex (d) at (0.5,0);
    % Outgoing operators
    \vertex (e) at (1.0, 0.8) 
    %{$O_{q_3}$};
    {$O_4$};
    \vertex (f) at (1.0, -0.8) 
    %{$O_{q_4}$};
    {$O_3$};
    % Diagram lines
    \diagram* {
      (a) -- (c) -- (d) -- (e),
      (b) -- (c) -- (d)--(f),
    };
    \node at (0, 0.25) 
    %{$O_{q_1 + q_2}$};
    {$O_s$};
  \end{feynman}
\end{tikzpicture}
\end{minipage}
\text{=  }
$\displaystyle \sum_{O_t}$
\begin{minipage}{0.35\columnwidth}
\begin{tikzpicture}
  \begin{feynman}
    % Incoming operators
    \vertex (a) at (-1.0, 0.8) 
    %{$O_{q_1}$};
    {$O_1$};
    \vertex (b) at (-1.0, -0.8) 
    %{$O_{q_2}$};
    {$O_2$};
    % Intermediate vertices
    \vertex (c) at (0, 0.4) ;
    \vertex (d) at (0, -0.4);
    % Outgoing operators
    \vertex (e) at (1.0, 0.8) 
    %{$O_{q_4}$};
    {$O_4$};
    \vertex (f) at (1.0, -0.8) 
    %{$O_{q_3}$};
    {$O_3$};
    % Diagram lines
    \diagram* {
      (a) -- (c) -- (e),
      (b) -- (d) -- (f),
      (c) -- (d)
    };
    \node[rotate=0] at (-0.25, 0) 
    %{$O_{q_2 + q_3}$};
    {$O_t$};
  \end{feynman}
\end{tikzpicture}
\end{minipage}
\caption{A schematic representation of the $s$-$t$ crossing equation.} 
\label{fig:Crossing}
\end{figure}

\textit{The conformal bootstrap} program ~\cite{Poland2018} has brought the study of higher-dimensional CFTs into the limelight with extensive work on both analytical and numerical fronts. The bootstrap philosophy attempts to solve the \emph{crossing symmetry} conditions coming from associativity of the OPE, with inputs from global symmetry and expected fusion rules for the operators. Crossing symmetry relates possible ways (or channels) of reducing  a four-point function $\langle \prod_{i=1}^4 O_i(\mathbf{r}_i) \rangle$ to two-point functions by replacing pairs of operators with their OPEs (see FIG. \ref{fig:Crossing}). The $s$-channel fusion ($1\to 2$, $3 \to 4$) and the $t$-channel fusion ($1\to 4$, $2 \to 3$) result in two expansions of the four-point function 
%that have overlapping regions of convergence 
and give the crossing equation 
%\begin{align}
%\label{eq:symbolicCrossing}
 $  \sum_{O_s} \lambda_{12}^{O_s}\lambda_{34}^{O_s} W_{O_s} = \sum_{O_t} \lambda_{14}^{O_t} \lambda_{23}^{O_t} W_{O_t}$. 
%\end{align}
The factors $W_O$ are fully de\-ter\-mined by conformal symmetry, while 
the CFT data $\{\Delta_i, \lambda_{ij}^k\}$ consisting of 
%the spectrum of 
scaling dimensions and OPE coefficients are to be found. Solutions $\{\Delta_i, \lambda_{ij}^k\}$ fully define consistent CFTs.
The $s$ and $t$ channels are obtained from each other by interchanges of indices of the operators (and their points of insertion): $s \leftrightarrow t \equiv 1\leftrightarrow 3$. Accordingly, starting with a function $f^{(s)} \equiv f(\mathbf{r}_1, \mathbf{r}_2; \mathbf{r}_3, \mathbf{r}_4)$ of four ordered arguments, we obtain, by permuting $1 \leftrightarrow 3$, another function $f^{(t)} \equiv f(\mathbf{r}_3, \mathbf{r}_2; \mathbf{r}_1, \mathbf{r}_4)$. Using this notation, we can write the four-point function as a product of a conformally covariant kinematic factor $\mathbb{K}_4^{(c)}$ and a {\it $G$ function} 
%(different in each channel)
\begin{align}
\langle \textstyle\prod_{i=1}^4 O_i(\mathbf{r}_i) \rangle 
= \mathbb{K}_4^{(s)} G^{(s)} = \mathbb{K}_4^{(t)} G^{(t)}.
\label{4-point-G-function}
\end{align}

The $G$ functions depend on the cross ratios 
\begin{align}
    u &= \frac{r_{12}^2 r_{34}^2}{r_{13}^2 r_{24}^2}, 
    &
    v &= \frac{r_{14}^2 r_{23}^2}{r_{13}^2 r_{24}^2},
    &
    \text{where}
    &&
    r_{ij} &= |\mathbf{r}_i - \mathbf{r}_j|,
    \label{eq:cross-ratios}
\end{align} 
and operator labels. The cross ratios get transformed upon crossing so that $G^{(s)} = G_{12,34}(u,v)$, $G^{(t)} = G_{32,14}(v,u)$. In terms of the $G$ functions, the $s$-$t$ crossing equation is
\begin{align}
\label{eq:stCrossing}
G^{(s)} u^{-\frac{\Delta_1 + \Delta_2}{2}} &= G^{(t)}  v^{-\frac{\Delta_2 + \Delta_3}{2}}.
\end{align}

Much of the bootstrap for\-ma\-lism is geared toward solving Eq.~\eqref{eq:stCrossing} self-consistently for unitary CFTs. Since any putative CFT for MF correlators contains infinitely many relevant operators and, thus, is nonunitary, we resort to novel, unorthodox methods that focus on the $G$ function and various physical inputs (similar 
%in spirit 
to the ``inverse bootstrap" method in~\cite{Li-InvBootstrap-2017}). We start by studying the Coulomb gas theories in the language of modern conformal bootstrap and use them as signposts to generalize the notion of Abelian fusion to higher dimensions. Then we show that the generalized Abelian fusion and crossing symmetry together yield a constraint on the spectrum of scaling dimensions in any $d$ that is analogous to the Vafa-Lewellen constraints~\cite{Vafa1988, Lewellen1988} known in CFT in $d=2$. Finally, additional physical assumptions specific to MF observables allow us to solve the constraint, leading to a quadratic dependence of the MF spectrum $\Delta_\gamma$ on $q_i$ in any dimension $d$. 

\textit{Coulomb gas theories with global conformal blocks.}---In $d = 2$, the Coulomb gas theories arise out of breaking the U(1) symmetry of the free boson $\phi$ by including a background charge $Q$ in the action~\cite{YellowBook, Levy2018}. A Coulomb gas CFT can be defined~\cite{Levy2018, Kislev-Odd-dimensional-2022} in any dimension $d \in \mathbb{N}$ by considering an action with a possibly nonlocal kinetic term $\propto\phi(-\square)^{\frac{d}{2}}\phi$. Such CFTs also arise as limits of \textit{generalized free fields}, where the scaling dimension of the field $\phi$ is tuned to $\Delta_\phi = 0$. In this limit, $\langle\phi\phi\rangle$ is logarithmic in any dimension which allows us to study vertex operators $V_{\alpha} \sim e^{d\alpha\phi}$. Following the conventions in~\cite{Levy2018, Kislev-Odd-dimensional-2022}, the scaling dimension of $V_\alpha$ is $\Delta_\alpha = d\alpha(Q - \alpha)$, and the multipoint functions satisfying the charge neutrality $\sum_i {\alpha_i} = Q$ are $\Big\langle \prod 
    V_{\alpha_i}(\mathbf{r}_i) \Big\rangle = \prod_{i<j} r_{ij}^{-2d\alpha_i\alpha_j}$.

Next, we derive the OPE of vertex operators in terms of primaries of the global conformal group by studying the conformal block expansion. Consider a four-point function of vertex operators which can be written in the form~\eqref{4-point-G-function} with the $G$ function 
\begin{align}
\label{eq:CG-Gfunction}
    G^{(s)}_\text{CG} 
    & = u^{\frac{1}{2} \Delta_{\alpha_1 + \alpha_2}} 
v^{\frac{1}{2}(\Delta_{\alpha_2 + \alpha_3} - \Delta_{\alpha_2} - \Delta_{\alpha_3})}.
\end{align}
This function is explicitly crossing symmetric [satisfies Eq.~\eqref{eq:stCrossing}] and has a convergent conformal block ex\-pan\-si\-on~\cite{Poland2018} in the $s$ channel in any dimension $d$,
\begin{align}
\label{eq:GlobalExpansion}
   G^{(s)}_\text{CG} = \textstyle\sum_{O} \lambda_{12}^O \lambda_{34}^O \, g_{\Delta_O, l_O}(u,v).
\end{align}
The conformal blocks $g_{\Delta_O, l_O}$ are often written as functions of $(z,\bar{z})$ related to the cross ratios $(u,v)$ by
\begin{align}
    u &= z\bar{z},
    &
    v &= (1 - z)(1 - \bar{z}).
\end{align}

In the $s$-channel limit, $r_{12} \approx r_{34} \ll r_{13} \approx r_{24} \approx r_{23} \approx r_{14}$, and thus, $u \rightarrow 0$, $v \rightarrow 1$; see Eq~\eqref{eq:cross-ratios}. Then, $z, \zbar \to 0$, and the $G$ function~\eqref{eq:CG-Gfunction} has the form 
\begin{align}
    G^{(s)} &= (z\zbar)^{\frac{\Delta^{(s)}}{2}} f(z, \zbar),
    \label{eq:G-z-zbar-one-family}
\end{align} 
where $f(z, \zbar)$ is a Taylor series symmetric in $(z, \zbar)$, and $\Delta^{(s)} = \Delta_{\alpha_1 + \alpha_2}$. In the Supplemental material~\cite{Supplementary} and Ref. \cite{GithubRepo}, we use the leading behavior of the conformal blocks in the $s$ channel~\cite{Dolan2000, Hogervorst2013} to show that any $G$ function of the form~\eqref{eq:G-z-zbar-one-family} admits the conformal block expansion
\begin{align}
    G^{(s)} = \textstyle\sum_{n,l \ge 0} \mu^{(n, l)} g_{\Delta^{(s)} + 2n+l, l}(z, \zbar)
    \label{eq:G-blocks-one-family}
\end{align}
in arbitrary dimensions $d \ge 2$. Conversely, any $G$ function that can be expanded as in Eq.~\eqref{eq:G-blocks-one-family} can also be written in the form of Eq.~\eqref{eq:G-z-zbar-one-family}.  

Let us denote global primaries as $[\tau, l]$ specifying their \emph{twist} $\tau \equiv \Delta - l$ and spin $l$. Then we say that the expansion~\eqref{eq:G-blocks-one-family} contains just one {\it twist family}~\cite{Li-InvBootstrap-2017}  consisting of the leading primary $[\Delta^{(s)}, 0]$ and subleading operators $[\Delta^{(s)} + 2n, l]$ which are constructed from its derivatives. The superscript of the product of the OPE coefficients $\mu^{(n, l)} \equiv \lambda_{12}^{(n,l)} \lambda_{34}^{(n,l)}$ identifies the operator $[\Delta^{(s)} + 2n, l]$ in the twist family.

Expanding the Coulomb gas $G$ function~\eqref{eq:CG-Gfunction} in global conformal blocks gives the OPE of $V_{\alpha_1}\times V_{\alpha_2}$ as~\cite{Supplementary} 
\begin{align}
&  [\Delta_{\alpha_1}, 0] \times [\Delta_{\alpha_2}, 0] \sim
 \sum_{n, l \ge 0}
    \lambda^{(n,l)}_{12}
[\Delta_{\alpha_1 + \alpha_2} + 2n, l],
    \label{eq:Coulomb-OPE}
\end{align}
where $n, l$ are non-negative integers, and the $(n, l)= (0,1)$ term is absent in the OPE. For two identical operators ($\alpha_1 = \alpha_2$), their OPE is completely specified by the conformal block expansion since we can extract the squared OPE coefficients (see~\cite{Supplementary} for explicit expressions in the $d = 2$ and $d = 4$ cases).

\textit{Generalized Abelian fusion.}---In the strict sense, Abelian fusion~\eqref{Abelian-fusion} cannot hold in CFTs in $d>2$, since an OPE written with finitely many \textit{global} conformal primaries cannot satisfy crossing~\cite{Rattazzi2008, Simmons-Duffin-Analytic-Bootstrap, Fitzpatrick2012, Alday-Zhiboedov}. Thus, we need to ge\-ne\-ra\-lize the notion of Abelian fusion to $d > 2$. Global conformal block ex\-pan\-sions of Coulomb gas correlators exhibit certain features that we adopt as the \emph{definition} of Abelian fusion in $d>2$: (1) All primary MF operators can be grouped into twist families, and (2) the OPE of any two leading MF primaries contains only one twist family:
    \begin{align}
        [\Delta_1, 0]\times[\Delta_2, 0] 
        \sim \textstyle\sum_{n, l \ge 0} 
       \lambda_{12}^{(n,l)} [\Delta + 2n, l].
        \label{eq:Abelian-fusion-d}
    \end{align}
%This definition is more general than Eq.~\eqref{eq:Coulomb-OPE}, where the three operators involved, $V_{\alpha_1}$, $V_{\alpha_2}$, an $V_{\alpha_1 + \alpha_2}$, are labeled by continuous parameters that add upon fusion. 

The generalized Abelian fusion~\eqref{eq:Abelian-fusion-d} and the related conformal block expansion~\eqref{eq:G-blocks-one-family} constrain a general scalar four-point $G$ function to have the form~\eqref{eq:G-z-zbar-one-family}. Since $z \zbar = u$ and $z + \zbar = u + 1 - v$ form a basis in the ring of symmetric functions of $z$ and $\zbar$, Eq.~\eqref{eq:G-z-zbar-one-family} can also be written as
\begin{equation}
    G^{(s)} = u^{\frac{\Delta^{(s)}}{2}} \textstyle\sum_{n \ge 0} f^{(s)}_n(v) u^{n}.
\end{equation}
%All powers of $u$ in this expansion differ from the leading power $\Delta^{(s)}/2$ by non-negative integers $n$, consistent with the even integer gaps between the spin-zero conformal blocks in Eq.~\eqref{eq:G-blocks-one-family}. 
The functions $f_n$ are arbitrary so far, and quantities with the superscript $(s)$ depend on the external dimensions in a channel-covariant manner. At this point, one can make a further simplification by assuming that the functions $f_r(v)$ can be represented as (possibly infinite) sums of power laws in $v$, i.e.,
\begin{equation}
    G^{(s)} = u^{\frac{\Delta^{(s)}}{2}} \textstyle\sum_{n,m \ge 0} C_{nm} v^{\sigma^{(s)}_m} u^{n}
    \label{eq:GfunctionAssumed}
\end{equation}
where $\sigma_m$ are unrelated real numbers and the coefficients $C_{nm}$ do not depend on the cross ratios $(u,v)$ or the external dimensions. The Coulomb gas theories are of this form with a single term; see Eq.~\eqref{eq:CG-Gfunction}. The generalized free field correlators~\cite{Fitzpatrick2012} with $\Delta_{\phi} \ge 0$ are similarly composed of sums of power laws in $u$ and $v$, although they do not satisfy Abelian fusion (integer gaps in powers of $u$). The above ansatz appeared for the case of a correlator of identical operators in Ref.~\cite{Li-InvBootstrap-2017}, which discussed the idea of building crossing-sym\-metric $G$ functions. Similarly, the authors of Ref.~\cite{Alday-Zhiboedov} use a version where the coefficients $C_{nm}$ are functions of $\log u, \log v$ (the logarithms come from anomalous dimensions of the subleading operators in the twist family). As our definition of Abelian fusion, Eq.~\eqref{eq:Abelian-fusion-d}, exactly fixes the dimensions of all subleading operators, the logarithms are unnecessary in our treatment. 

\textit{Constraints on the $G$ function from crossing.} 
%Now we explore constraints on the Abelian $G$ function~\eqref{eq:GfunctionAssumed} imposed by crossing symmetry. 
Substituting the Abelian $G$ function~\eqref{eq:GfunctionAssumed}
%this function 
into 
%the crossing equation
Eq.~\eqref{eq:stCrossing}, we obtain an equation
%\begin{align}
%& u^{\frac{\Delta^{(s)}}{2} - \frac{\Delta_1 + \Delta_2}{2}} %\textstyle\sum_{n,m \ge 0} C_{nm} v^{\sigma^{(s)}_m} u^{n}
%\nonumber \\
%& = v^{\frac{\Delta^{(t)}}{2} - \frac{\Delta_2 + \Delta_3}{2}} %\textstyle\sum_{n,m \ge 0} C_{nm} u^{\sigma^{(t)}_m} v^{n}.
%\label{eq:s-t-crossing}
%\end{align}
%This equation 
that enforces a structure on the $G$ function 
%that is 
understandable in terms of crossing symmetric building blocks~\cite{Li-InvBootstrap-2017}, such that any truncation up to $(N, M)$ of the double sum in Eq.~\eqref{eq:GfunctionAssumed} is also crossing symmetric; see the Supplemental material~\cite{Supplementary} for details. 
%In this case all terms in Eq.~\eqref{eq:GfunctionAssumed} fall into two groups: single terms that are crossing-symmetric by themselves, and pairs of terms that get exchanged under crossing. 
The result is the $G$ function
\begin{align}
\label{eq:ChannelDependentG}
G^{(s)}(u,v) &= u^{\frac{\Delta^{(s)}}{2}}v^{\frac{\Delta^{(t)}}{2} - \frac{\Delta_2 + \Delta_3}{2}}\Big(\textstyle\sum_{k\ge 0}S_k(uv)^k 
\nonumber \\
& \quad + \textstyle\sum_{j\ge 1, k \ge 0}D_{jk}(uv)^k(u^j + v^j)\Big), 
\end{align}
where we use $S_k$ and $D_{jk}$ to represent the coefficients of the crossing-symmetric terms and pairs, respectively. Thus, we adopt  Eq.~\eqref{eq:ChannelDependentG} as the generic form of the Abelian $G$ function~\eqref{eq:GfunctionAssumed} that \textit{also} satisfies $s$-$t$ crossing.

\textit{Excluding the spin-1 operator.}---Focusing on the last part of the puzzle, we expand the $G$ function~\eqref{eq:ChannelDependentG} in conformal blocks in arbitrary dimensions to first few orders in $z$ and $\zbar$. By construction, the first block that appears in the expansion is $[\Delta^{(s)}, 0]$. The product of the OPE coefficients of the leading block is read off as $\mu^{(0,0)} \equiv S_0$. The coefficient $\mu^{(0,1)}$ of the spin-1 block $[\Delta^{(s)}, 1]$ can be obtained by matching the coefficients of the series for the order $\sim(z\zbar)^{\Delta^{(s)}/2}(z + \zbar)$ as
\begin{align}
\label{eq:spin1OPECoeff}
\frac{\mu^{(0,1)}}{S_0}
&= \frac{\Delta_2 + \Delta_3 - \Delta^{(t)}}{2}(1 + \mathcal{P}) - \mathcal{Q}
\nonumber \\
& 
- \frac{(\Delta^{(s)} - \Delta_{1} + \Delta_2)(\Delta^{(s)} + \Delta_{3} - \Delta_4)}{4\Delta^{(s)}},
\end{align}
where the sums 
%\begin{align}
    $\mathcal{P} \equiv \sum_{j\ge 1} 
    D_{j0}/S_0$, 
    $\mathcal{Q} \equiv \sum_{j\ge 1} 
    jD_{j0}/S_0$
%\end{align}
must converge for the OPE coefficient to be well defined.

This spin-1 operator cannot appear in any OPE of two \textit{identical} scalar operators on general grounds. Indeed, $O(x_1) O(x_2)$ is even with respect to the interchange $x_1 \leftrightarrow x_2$ but a spin-1 operator must appear in the OPE as $ \sim(x_1 - x_2) \cdot \partial O_{\Delta^{(s)}}((x_1 + x_2)/2)$ which is odd. Exploiting this fact, we set $O_{2} \equiv O_1$ in which case $\mu^{(0,1)} = 0$, and Eq.~\eqref{eq:spin1OPECoeff} becomes a constraint on $\Delta$'s:
\begin{align}
\label{eq:DeltaConstraint}
\Delta^{(s)} + \Delta_3 - \Delta_4+ 4\mathcal{Q} 
= 2(\Delta_1 + \Delta_3 - \Delta^{(t)})(1 + \mathcal{P}).
\end{align}

In the context of MF correlators at ATs, we identify $\Delta^{(s)} \equiv \Delta_{\gamma_1 + \gamma_2}$, $\Delta^{(t)} \equiv \Delta_{\gamma_3 + \gamma_2}$. The neutrality condition $\sum_i \gamma_i = -\rho_b$ fixes $\Delta_4 = \Delta_{-\rho_b - \gamma_1 - \gamma_2 - \gamma_3} = \Delta_{\gamma_1 + \gamma_2 + \gamma_3}$. %Furthermore, 
Now the continuity of MF spectra allows us to choose $\gamma_1 = \gamma_2 = \epsilon e_i$, where $e_i = (0,\ldots, 1, \ldots, 0)$ (unit in the $i$th place), with $\epsilon \ll 1$, and $\gamma_3 = \gamma$ in Eq.~\eqref{eq:DeltaConstraint}.
%, which gives
%\begin{align}
%\label{eq:DeltaConstraint-eps}
%&\Delta_{2\epsilon e_i} + \Delta_\gamma - \Delta_{\gamma + 2\epsilon %e_i} + 4\mathcal{Q} 
%\nonumber \\
%& \quad = 2(\Delta_{\epsilon e_i} + \Delta_\gamma - \Delta_{\gamma + %\epsilon e_i})(1 + \mathcal{P}).
%\end{align}
%Now 
Then we can expand in orders of $\epsilon$~\cite{Supplementary}
%(see the Supplementary material~\cite{Supplementary}), 
which gives $\mathcal{Q} = \mathcal{P} = 0$, and our main result:
\begin{quote}
The only MF spectrum $\Delta_\gamma$ which satisfies generalized Abelian fusion and crossing symmetry has the form given in Eq.~\eqref{parabolic-spectrum}.
\end{quote}
Going back to Eq.~\eqref{eq:DeltaConstraint}, we substitute $\mathcal{P} = \mathcal{Q} = 0$, and the quadratic solution for $\Delta_\gamma$ to find that the constraint
\begin{equation}
\label{eq:DeltaConstraint-1}
2\Delta_{\gamma_1} + \Delta_{\gamma_3} - 2\Delta_{\gamma_1 + \gamma_3} - \Delta_{2\gamma_1} + \Delta_{2\gamma_1 + \gamma_3} = 0
\end{equation}
correctly picks out Abelian CFTs in $d \ge 2$, and thus is the appropriate generalization of the 2D Vafa-Lewellen constraint with a single exchanged Virasoro primary.

\textit{Summary and outlook.}---Using conformal invariance, we have shown that any Abelian CFT in $d > 2$ must be intimately related to the Coulomb gas theory, and have a quadratic spectrum. Our main assumptions, fundamentally related to each other, were the Abelian fusion~\eqref{eq:Abelian-fusion-d} and the form~\eqref{eq:GfunctionAssumed} for the $G$ function.
%To summarize, the set of assumptions used to prove the exact parabolicity of the MF spectrum in $d \ge 2$ is the following:
%\begin{enumerate}
%    \item The MF operators $O_\gamma$ are global primaries in some CFT, so that their correlators solve the crossing equations.
%    \item The fusion of two MF operators is of the general form~\eqref{eq:Abelian-fusion-d} (generalized Abelian fusion).
%    \item The $G$-functions of the charge-neutral 4-point cor\-re\-la\-tors of MF operators admit an expansion of the form $G^{(s)} = \sum_{n,m} C_{nm}u^{\Delta_{\gamma_1 + \gamma_2}/2 + n}v^{\sigma_m}$.
%\end{enumerate}
%The second and third assumptions are our strongest, and are fundamentally related to each other. 
As in the case of weakly perturbed CFTs~\cite{Alday-Zhiboedov}, it remains to be seen if the generalized Abelian CFT defined here could be perturbed so that the derivative operators gain anomalous dimensions.

Let us discuss the implication 
%of the first assumption, that 
of conformal invariance. As we have summarised earlier, perturbative analytical results in $d = 2 + \epsilon$ and numerical simulations in $d = 3,4,5$ have shown that the MF spectra for generic ATs are in fact, \textit{not} parabolic~\cite{Obuse-Boundary-2008, Evers-Multifractality-2008, Puschmann-Quartic-2021, Karcher2021, Karcher2022, Karcher2022b, Karcher-Generalized-2023, Rodriguez-Multifractal-2011, Ujfalusi-Finite-size-2015, Lindinger-Multifractal-2017, Tarquini-Critical-2017}. In light of our result, it follows that conformal invariance is likely lost at ATs. The alternative scenario advocated in Refs.~\cite{Zirnbauer-The-integer-2019, Arenz-Wegner-2023} is that the symmetries of the sigma models that were used to derive Abelian fusion and Weyl symmetry are spontaneously broken at the critical point. We believe this alternative to be unlikely, since it contradicts the vast body of literature on ATs, including the aforementioned numerical confirmations of the Weyl symmetry~\cite{Evers-Multifractality-2003, Mirlin-Wavefunction-2003, Mirlin-Exact-2006, Mildenberger-Boundary-2007, Mildenberger-Wave-2007, Obuse-Multifractality-2007, Obuse-Boundary-2008, Evers-Multifractality-2008, Vasquez-Multifractal-2008, Rodriguez-Multifractal-2008, Rodriguez-Multifractal-2011, Karcher2021, Karcher2022, Karcher2022b, Karcher-Generalized-2023}. Thus, we propose ATs as examples of systems where scale invariance \textit{does not} imply conformal invariance.  

Perturbative MF spectra at random critical points~\cite{Ludwig-Infinite-1990, Monthus-Symmetry-2009} are also nonparabolic, suggesting lack of conformal invariance. Moreover, the authors of Ref.~\cite{Hastings-Breakdown-2001} argued that conformal invariance generically breaks down at strongly random fixed points. On the other hand, most systems where the MF spectrum is known to be parabolic, are also conformally invariant~\footnote{The parabolicity of MF spectra in the context of fully developed turbulence~\cite{Benzi-Multifractal-2022} is not a settled issue. Since there are no tools to verify Abelian fusion, we cannot apply our results to talk about conformal invariance~\cite{Falkovich-Operator-2015, Oz-On-scale-2018}.}. These include 2D Dirac fermions in random gauge potentials~\cite{Ludwig-Integer-1994, Castillo-Exact-1997, Nersesyan-Disorder-1995, Chamon-Localization-1996, Mudry-Two-dimensional-1996, Caux-Disordered-1998, Caux-Exact-1998}, a recent proposal for the critical-point theory of the integer quantum Hall transition~\cite{Bondesan2016, Zirnbauer-The-integer-2019, Zirnbauer-Marginal-2021}, Coulomb gas and Liouville CFTs in arbitrary dimensions~\cite{Levy2018, Kislev-Odd-dimensional-2022}, and rigorous probabilistic studies of 2D quantum gravity and Liouville CFT~\cite{Rhodes-Gaussian-2013, Rhodes-Lecture-2016, Vargas-Lecture-2017}. All of these results support the picture where parabolicity of MF spectra and conformal invariance go hand in hand, and that both are absent at critical points in random and disordered systems~\footnote{However, we cannot exclude a logically possible situation where the MF spectrum is parabolic but there is no conformal invariance.}. 

A natural extension of our Letter is to consider implications of conformal invariance for multifractality near boundaries of finite systems~\cite{Subramaniam-Surface-2006, Mildenberger-Boundary-2007, Obuse-Multifractality-2007, Obuse-Boundary-2008, Evers-Multifractality-2008, Obuse-Corner-2008, Subramaniam-Boundary-2008, Obuse-Conformal-2010, Babkin-Generalized-2023}  using crossing symmetry and conformal bootstrap in a boundary CFT. 

%Going forward, we believe that it is important to better understand the relationship between scale and conformal invariance, especially for non-unitary theories. ATs and MF correlators in general may provide a fertile ground for such understanding. To this end, it would be interesting to verify the validity of Abelian fusion for generic MFs outside the context of ATs. Then our result of parabolic MF spectra would apply, functioning as a test of conformal invariance for the MF correlators. 

 We thank Yan Fyodorov, Dalimil Maz\'a\v{c}, Marco Meineri, Alexander Mirlin, Yaron Oz, Lorenzo Quintavalle, Sylvain Ribault, Shivaji Sondhi, and Bernardo Zan  for useful discussions. This research was supported by Grant No.~2020193 from the United States-Israel Binational Science Foundation.

\bibliography{bibliography}% Produces the bibliography via BibTeX.
\clearpage

\onecolumngrid
\section*{Supplementary material}
\pagenumbering{roman}
\renewcommand{\theequation}{S.\arabic{equation}}
\setcounter{equation}{0}
\subsection{Subleading behavior of global conformal blocks}
\label{sec:subleadingBlock}

Start with the Casimir equation solved by the spin-0 conformal blocks
\begin{equation}
    \mathcal{D}g_{\Delta, 0}^{\Delta_{12}, \Delta_{34}}(z, \zbar) = C_{\Delta, 0}g_{\Delta, 0}^{\Delta_{12}, \Delta_{34}}(z, \zbar)
\end{equation}
where $\mathcal{D}$ is the Casimir operator of $SO(d+1,1)$:
\begin{align}
    \mathcal{D} &= \mathcal{D}_z + \mathcal{D}_{\zbar} + 2(d-2)\frac{z\zbar}{z - \zbar}\left[(1-z)\partial_z - (1-\zbar)\partial_{\zbar}\right], 
    \\
    \mathcal{D}_z &= 2z^2(1-z)\partial_z^2 - (2 + \Delta_{34} - \Delta_{12})z^2\partial_z + \frac{\Delta_{12}\Delta_{34}}{2}z, 
\end{align}
and $C_{\Delta, 0} \equiv \Delta(\Delta - d)$ is the eigenvalue of the Casimir operator. Based on the leading form of $g_{\Delta, 0}$ and the symmetricity with respect to $z$ and $\zbar$, we use the following power series as an ansatz for the solution:
\begin{equation}
    g_{\Delta, 0}(z, \zbar) = (z\zbar)^{\alpha}\sum_{r,s = 0}^{\infty}\frac{\kappa_{rs}}{1 + \delta_{rs}}(z^r\zbar^s + z^s\zbar^r).
\end{equation}
For the subleading behavior of the block, we need to compute the coefficient $\kappa_{10}$. Substituting the ansatz into the Casimir equation and extracting the terms at leading order gives 
\begin{equation}
    C_{\Delta, 0} = \Delta(\Delta - d) = 4\alpha(\alpha - 1) - 2(d - 2)\alpha
\end{equation}
which is solved by $\alpha = \Delta/2$ as expected. $\kappa_{00}$ is a normalization that we can set to unity. Now considering the subleading order terms, we have
\begin{align}
    2\kappa_{10}\bigg((\alpha+1)_2 - (\alpha)_2 - (d-2)\alpha\bigg) - 2(\alpha)_2 - \alpha(2 + \Delta_{34} - \Delta_{12}) + \frac{\Delta_{12}\Delta_{34}}{2} = \kappa_{10}\Delta(\Delta - d).
\end{align}
$\kappa_{10}$ is read off from the equation as
\begin{equation}
    \kappa_{10} = \frac{(\Delta - \Delta_{12})(\Delta + \Delta_{34})}{4\Delta}
\end{equation}
with no dependence on dimension $d$.

\subsection{Global conformal block decomposition of Coulomb gas-like correlator}
\label{sec:GeneralGlobalBlockExpansion}
Consider a $G$-function which admits a series expansion at $z, \zbar \rightarrow 0$ of the form
\begin{equation}
\label{eq:genericGfunction}
    G(z,\zbar) = (z\zbar)^p\sum_{i,j = 0}^{\infty} \Cee_{ij} z^i\zbar^j
\end{equation}
with $\Cee_{ij} = \Cee_{ji}$. The Coulomb gas correlator (Eq. 10 in the main text) is a particular example, with 
\begin{equation}
    2p = \Delta_{\alpha_1 + \alpha_2} \hspace{0.5in} \text{and} \hspace{0.5in} \Cee_{ij} = \binom{i + d\alpha_2\alpha_3 - 1}{i}\binom{j + d\alpha_2\alpha_3 - 1}{j}.
\end{equation}

Let us show that the given $G(z, \zbar)$ can be expanded in terms of global conformal blocks in any dimension $d$ with a single double trace family $[2p + 2n, l]$. In other words, we show that 
\begin{equation}
\label{eq:genericExpansion}
    G(z, \zbar) = \sum_{n, l \ge 0}\mu^{(n, l)}g_{2p + 2n + l, l}(z, \zbar)
\end{equation}
where $\mu^{(n,l)}$ is a product of two OPE coefficients. The proof will be based upon the idea that the expansion Eq. \ref{eq:genericExpansion} can be uniquely fixed order-by-order in powers of $z$ and $\zbar$.

As noted in the main text, The leading order behavior of the global conformal blocks in any dimensions is
\begin{equation}
\label{eq:BlockLeading}
    g_{\Delta , l} \sim \mathcal{N}_{d, l}(z\zbar)^{\Delta/2}\text{Geg}_l^{d/2 - 1}\left(\frac{z + \zbar}{2\sqrt{z\zbar}}\right)
\end{equation}
where we work with the normalization
\begin{equation}
    \mathcal{N}_{d, l} = \frac{l!}{\left( d/2 - 1\right)_l}.
\end{equation}
Focusing on the double-trace blocks $g_{2p + 2n + l, l}$, we know further that the subleading terms should be expandable as a power series
\begin{equation}
\label{eq:GlobalBlockPowerSeries}
    g_{\Delta + 2n + l, l}(z, \zbar) = (z\zbar)^{\Delta/2}\left(\mathcal{N}_{d, l}(z\zbar)^{l/2 + n}\text{Geg}_{l}^{d/2 - 1}\left(\frac{z + \zbar}{2\sqrt{z\zbar}}\right) + \sum_{r + s > (n + l/2)}\frac{\kappa^{(n, l)}_{rs}}{1 + \delta_{rs}}z^r\zbar^s\right)
\end{equation}
where $r, s$ are integers and $\kappa_{rs} = \kappa_{sr}$ \cite{Hogervorst2013}.

The Gegenbauer polynomial for $l = 0$ is just 1, so for the leading degree, the block $g_{2p, 0} \sim (z\zbar)^{p}$ has the correct power-law to match $G(z,\zbar) = \Cee_{00}(z\zbar)^p + \ldots$. 

Next, consider degree 1 terms in the series, $(z\zbar)^{p}(\Cee_{10}(z + \zbar) + \ldots)$. On the conformal blocks side, the subleading contribution from the leading block at this degree can be balanced by adding a new term $[2p, 1]$ with the correct power-law behavior,
\begin{equation}
    g_{2p + 1, 1} \sim (z\zbar)^{p + \frac{1}{2}}\left(\frac{z + \zbar}{2\sqrt{z\zbar}}\right). 
\end{equation}
The OPE coefficient is read off by equating the two series upto this order,
\begin{equation}
\label{eq:FirstOrderOPECoeff}
    \mu^{(1,1)} = \Cee_{10} - \mu^{(0,0)}\frac{\partial}{\partial z}\left.\frac{g_{2p, 0}(z, \zbar)}{(z\zbar)^p}\right|_{z = \zbar = 0} = \Cee_{10} - \Cee_{00}\frac{(2p - \Delta_{12})(2p + \Delta_{34})}{8p}
\end{equation}
where we have used the subleading coefficient of the spin-zero block derived in \ref{sec:subleadingBlock}.
Now, consider an arbitrary term in the $G$-function expansion $\sim \Cee_{ij} z^{i}\zbar^j$. There is always a conformal block in the double-trace family (in this case $g_{2p + i + j, i - j}$) with the correct leading behavior to fix the series at this order. Looking closely at Eq. \ref{eq:BlockLeading}, we have (with $j = n$ and $i - j = l$)
% \begin{align}
%     g_{2p + i + j, i-j } &\sim (z\zbar)^{p + (i+j)/2}\frac{(i-j)!}{(d/2 - 1)_{i-j}}\text{Geg}_{i-j}^{d/2 - 1}\left(\frac{z + \zbar}{2\sqrt{z \zbar}}\right) \\
%     &= (z\zbar)^{p + (i + j)/2}\frac{(i-j)!}{(d/2 - 1)_{i-j}} \sum_{k = 0}^{\lfloor\frac{i-j}{2}\rfloor}\frac{\left(d/2 - 1\right)_{i-j-k}}{k!(i-j-2k)!}(-1)^k\left(\frac{z + \zbar}{\sqrt{z\zbar}}\right)^{i-j-2k}\\
%     \label{eq:ExpandedConformalBlock}
%     &= (z\zbar)^p\sum_{k=0}^{\lfloor\frac{i-j}{2}\rfloor}\frac{(-1)^k(i-j-2k)_{2k}}{k!(d/2 + i - j- k - 1)_k}(z\zbar)^{j + k}(z + \zbar)^{i -j - 2k}.
% \end{align}

\begin{align}
    g_{2p + 2n + l, l } &\sim (z\zbar)^{p + n + l/2}\frac{l!}{(d/2 - 1)_l}\text{Geg}_l^{d/2 - 1}\left(\frac{z + \zbar}{2\sqrt{z \zbar}}\right) \\
    &= (z\zbar)^{p + n + l /2}\frac{l!}{(d/2 - 1)_{l}} \sum_{k = 0}^{\lfloor\frac{l}{2}\rfloor}\frac{\left(d/2 - 1\right)_{l-k}}{k!(l-2k)!}(-1)^k\left(\frac{z + \zbar}{\sqrt{z\zbar}}\right)^{l-2k}\\
    \label{eq:ExpandedConformalBlock}
    &= (z\zbar)^p\sum_{k=0}^{\lfloor\frac{l}{2}\rfloor}\frac{(-1)^k(l-2k)_{2k}}{k!(d/2 + l- k - 1)_k}(z\zbar)^{k + n}(z + \zbar)^{l - 2k}.
\end{align}

Setting $k = 0$ in Eq.~\eqref{eq:ExpandedConformalBlock} yields a term proportional to $(z\zbar)^p(z^i\zbar^j)$ with coefficient 1 as desired. There are additional terms even at the leading order in the conformal block, but all of them are of the same degree in $z, \zbar$: $i + j$. Therefore, for each degree, the terms in the series of the same degree must be fixed in \textit{descending order} in $l$, so that the $l = 0$ or $l = 1$ term is fixed last. For example, in the quadratic degree, there are two terms in the G-function series, $\Cee_{20}(z\zbar)^p(z^2 + \zbar^2)$ and $\Cee_{11}(z\zbar)^p(z\zbar)$. Of the two corresponding blocks $g_{2p + 2, 2}$ and $g_{2p + 2, 0}$, $g_{2p + 2,2} \sim (z\zbar)^{p}(z^2 + \zbar^2 + z\zbar)$ contributes to the $z\zbar$ term but the block $g_{2p + 2, 0} \sim (z\zbar)^{p+1}$ does not feed back into the former. 

In summary, the recursive process can be codified into a formula for calculating $\mu^{(n,l)}$ given $C_{ij}$ as
\begin{equation}
\label{eq:RecursiveBlockExpansion}
    \mu^{(j, i - j)} = \Cee_{ij} - \frac{1}{i!j!}\frac{\partial^i}{\partial z^i}\frac{\partial^i}{\partial \zbar^j}\left( \frac{1}{(z\zbar)^p}\left[ \sum_{2\alpha < (i + j)}\sum_{\beta = 0}^{i + j - 1 - 2\alpha} \mu^{(\alpha, \beta)}g_{2p + 2\alpha + \beta, \beta} + \sum_{\alpha = 0}^{j - 1}\mu^{(\alpha, i + j - 2\alpha)}g_{(2p + i + j, i + j - 2\alpha)}\right]\right)_{z = \zbar = 0}.
\end{equation}
Using the form of double-trace blocks in Eq. \ref{eq:GlobalBlockPowerSeries}, this becomes
\begin{equation}
\label{eq:ExpandedRecursiveBlockExpansion}
    \mu^{(j, i - j)} = \Cee_{ij} - \left[ \sum_{2\alpha < (i + j)}\sum_{\beta = 0}^{i + j - 1 - 2\alpha} \mu^{(\alpha, \beta)}\kappa^{(\alpha, \beta)}_{ij} + \sum_{\alpha = 0}^{j - 1}\mu^{(\alpha, i + j - 2\alpha)}\mathscr{K}_{\alpha}\right].
\end{equation}
where the numbers $\mathscr{K}_\alpha$ are given by
\begin{equation}
    \mathscr{K}_\alpha = \sum_{k = j - \alpha}^{\lfloor\frac{i + j}{2} - \alpha\rfloor} \binom{i+j - 2\alpha - 2k}{i - \alpha - k}\frac{(-1)^k(i + j - 2\alpha -2k)_{2k}}{k!(d/2 + i + j -2\alpha - k - 1)_k}.
\end{equation}
Of course, to calculate $\mu^{(j, i -j)}$ more explicitly one requires the coefficients $\kappa_{ij}$ in the definition of the conformal blocks, which is not available in generic dimensions. But we still have the result
\begin{quote}
    \textit{Any G-function of the form $G(z, \zbar) = (z\zbar)^pf(z,\zbar)$ where $f(z, \zbar)$ has a convergent Taylor series expansion at $z = \zbar = 0$ can be expanded in  conformal blocks from a single twist family $[2p + 2n, l]$.}
\end{quote}

\subsection{Coulomb gas OPE coefficients}

Following the procedure outlined in \ref{sec:GeneralGlobalBlockExpansion}, we performed the global conformal block expansion of the Coulomb Gas $G$-function in $d = 2$ and $d = 4$. In this section we present some interesting observations and the first few coefficients in the OPE of identical scalars. 

Consider the Coulomb Gas $G$-function reproduced here for convenience,
\begin{equation}
    G(u,v) = u^{\frac{d}{2}(\alpha_3 + \alpha_4)}v^{-d\alpha_2\alpha_3}.
\end{equation}
One sees that it can be rewritten as a series expansion of the kind discussed in \ref{sec:GeneralGlobalBlockExpansion}, 
\begin{equation}
    G(z, \zbar) = (z\zbar)^{\frac{\Delta_{\alpha_1 + \alpha_2}}{2}}\left(\sum_{i,j = 0}^{\infty}\binom{i + d\alpha_2\alpha_3 - 1}{i}\binom{j + d\alpha_2\alpha_3 - 1}{j}z^i\zbar^j\right)
\end{equation}
where we have used $\Delta_{\alpha_1 + \alpha_2} = d(\alpha_1 + \alpha_2)(\alpha_3 + \alpha_4)$ and charge neutrality. At leading order in any $d$, the coefficient $\Cee_{00} = 1$ in the series above. Thus the leading block in the expansion is $[\Delta_{\alpha_1 + \alpha_2}, 0]$ with coefficient $\mu^{(0,0)} = 1$. At the next degree, the only block to be considered is $[\Delta_{\alpha_1 + \alpha_2} + 1, 1]$, with the product OPE coefficient given by Eq. \ref{eq:FirstOrderOPECoeff}, which turns out to be 
\begin{equation}
    \mu^{(0,1)} = \binom{d\alpha_2\alpha_3}{1}  - \frac{(\Delta_{\alpha_1 + \alpha_2} - \Delta_{12})(\Delta_{\alpha_1 + \alpha_2} + \Delta_{34})}{4\Delta_{\alpha_1 + \alpha_2}}.
\end{equation}
As the discussion pans out in the key argument of the main text, we notice  upon simplification that generically, for all the Coulomb Gas theories,
\begin{equation}
    \mu^{(0,1)} = 0.
\end{equation}
Let us stress that for $d = 2$, it is expected that this OPE coefficient vanishes in the global block expansion. It is a manifestation of the result that in a Verma module, there are \textit{no quasiprimaries} at level 1. But as we have shown, this fact explicitly generalizes to higher dimensions for Coulomb Gas theories. 

Computing the expansion to higher degrees in $d = 2$, we observe that the twist-two $(n = 1)$ family of operators $[\Delta_{\alpha_1 + \alpha_2} + 2, l]$ is missing from the OPE (their OPE coefficients vanish for any $G$-function independent of $\alpha_i$). Additionally, as we expect, in the OPE of identical operators, the coefficients of odd-spin operators vanish. For example, 
\begin{equation}
    \mu^{(0,3)} = \frac{2\alpha_1\alpha_2\alpha_3\alpha_4(\alpha_1 - \alpha_2)(\alpha_3 - \alpha_4)}{3(1 + (\alpha_1 + \alpha_2)(\alpha_3 + \alpha_4))(2 + (\alpha_1 + \alpha_2)(\alpha_3 + \alpha_4))}
\end{equation}
and we see that $\mu^{(0,3)} = 0$ if $\alpha_1 = \alpha_2$ or $\alpha_3 = \alpha_4$. The non-zero coefficients in the OPE of two identical scalars are tabulated in Table~\ref{tab:OPE}, left column. (For the case of non-identical scalars, see~\cite{GithubRepo}).

\begin{table}[h]
%   \begin{center}
     \begin{tabular}{c | c}
    \toprule
     $(n,l)$ &  $(\lambda^{(n,l)})^2$ \\ %[0.5ex]
    \midrule
     $(0,0)$  & 1\\
     \midrule
     $(0,2)$  & $\dfrac{2\alpha^4}{8\alpha^2 + 1}$ \\
     \midrule
     $(0,4)$  & $\dfrac{\alpha^4(2\alpha^2 +1)^2}{2(64\alpha^4 + 64\alpha^2 + 15)}$  \\
     \midrule
     $(2,0)$  & $\dfrac{4\alpha^8}{(8\alpha^2 + 1)^2}$  \\
     \midrule
     $(0,6)$  & $\dfrac{\alpha^4(2\alpha^4 + 3\alpha^2 + 1)^2}{3(8\alpha^2 + 5)(8\alpha^2 + 7)(8\alpha^2 + 9)}$   \\
     \midrule
     $(2,2)$  & $\dfrac{\alpha^8(2\alpha^2 + 1)^2}{(8\alpha^2 + 1)(8\alpha^2 + 3)(8\alpha^2 + 5)}$ \\
     \midrule
     $(0,8)$  & $\dfrac{\alpha^4(4\alpha^6 + 12\alpha^4 + 11\alpha^2 + 3)^2}{24(8\alpha^2 + 7)(8\alpha^2 + 9)(8\alpha^2 + 11)(8\alpha^2 + 13)}$\\
     \midrule
     $(2,4)$  & $\dfrac{2\alpha^8(2\alpha^4 + 3\alpha^2 + 1)^2}{3(8\alpha^2 + 1)(8\alpha^2 + 5)(8\alpha^2 + 7)(8\alpha^2 + 9)}$\\
     \midrule
     $(4,0)$  & $\dfrac{\alpha^8(2\alpha^2 + 1)^4}{4(64\alpha^4 + 64\alpha^2 + 15)^2}$\\
     \midrule
     \bottomrule
    \end{tabular}
    \hspace{2cm}
    \begin{tabular}{c | c}
    \toprule
     $(n,l)$ & $(\lambda^{(n,l)})^2$\\ %[0.5ex]
    \midrule
     $(0,0)$  & 1\\
     \midrule
     $(0,2)$  & $\dfrac{8\alpha^4}{16\alpha^2 + 1}$ \\
     \midrule
     $(1,0)$  & $\dfrac{8\alpha^4}{1-16\alpha^2}$  \\
     \midrule
     $(0,4)$  & $\dfrac{2\alpha^4(4\alpha^2 + 1)^2}{(16\alpha^2+3)(16\alpha^2 + 5)}$  \\
     \midrule
     $(2,2)$  & $-\dfrac{2\alpha^4(4\alpha^2 + 1)^2}{3 + 64(4\alpha^4 + \alpha^2)}$   \\
     \midrule
     $(2,0)$  & $\dfrac{64\alpha^8}{256\alpha^4 - 1}$ \\
     \midrule
     $(0,6)$  & $\dfrac{4\alpha^4(8\alpha^4 + 6\alpha^2 +1)^2}{3(16\alpha^2+5)(16\alpha^2 + 7)(16\alpha^2+9)}$\\
     \midrule
     $(1,4)$  & $-\dfrac{4\alpha^4(8\alpha^4 + 6\alpha^2 + 1)^2}{3(16\alpha^2 + 3)(16\alpha^2 + 5)(16\alpha^2 + 7)}$\\
     \midrule
     $(2,2)$  & $\dfrac{16\alpha^8(4\alpha^2 + 1)^2}{(16\alpha^2 - 1)(16\alpha^2+3)(16\alpha^2 + 5)}$\\
     \midrule
     $(3,0)$  & $-\dfrac{16\alpha^8(4\alpha^2 + 1)^2}{(16\alpha^2 + 1)^2(16\alpha^2 + 3)}$\\
     \bottomrule
    \end{tabular}
%    \end{center}
    \caption{Nonzero OPE coefficients for the coulomb gas correlator $\langle V_\alpha V_\alpha V_\alpha V_\alpha\rangle$ in the $s$-channel with $\alpha \equiv Q/4$ to satisify charge neutrality. Left: The OPE coefficients up to degree 8 
%    $(n = 8)$ 
    in $d = 2$ 
    Right: The OPE coefficients up to degree 6 in $d = 4$.}
%    \label{tab:OPE2d}
    \label{tab:OPE}
\end{table}

Similarly, for $d = 4$, we obtain the global conformal block expansion of the $G$-function. We notice that the odd spin OPE coefficients still vanish, the twist-two family is no longer missing in the OPE. We also see that not all OPE coefficient squares are positive definite, owing to the non-unitary nature of the theory. The coefficients are tabulated in Table~\ref{tab:OPE}, right column.

\begin{comment}
\begin{table}[h]
    \begin{center}
     \begin{tabular}{c | c}
    \toprule
     $(n,l)$ & $(\lambda^{(n,l)})^2$\\ %[0.5ex]
    \midrule
     $(0,0)$  & 1\\
     \midrule
     $(0,2)$  & $\dfrac{8\alpha^4}{16\alpha^2 + 1}$ \\
     \midrule
     $(1,0)$  & $\dfrac{8\alpha^4}{1-16\alpha^2}$  \\
     \midrule
     $(0,4)$  & $\dfrac{2\alpha^4(4\alpha^2 + 1)^2}{(16\alpha^2+3)(16\alpha^2 + 5)}$  \\
     \midrule
     $(2,2)$  & $-\dfrac{2\alpha^4(4\alpha^2 + 1)^2}{3 + 64(4\alpha^4 + \alpha^2)}$   \\
     \midrule
     $(2,0)$  & $\dfrac{64\alpha^8}{256\alpha^4 - 1}$ \\
     \midrule
     $(0,6)$  & $\dfrac{4\alpha^4(8\alpha^4 + 6\alpha^2 +1)^2}{3(16\alpha^2+5)(16\alpha^2 + 7)(16\alpha^2+9)}$\\
     \midrule
     $(1,4)$  & $-\dfrac{4\alpha^4(8\alpha^4 + 6\alpha^2 + 1)^2}{3(16\alpha^2 + 3)(16\alpha^2 + 5)(16\alpha^2 + 7)}$\\
     \midrule
     $(2,2)$  & $\dfrac{16\alpha^8(4\alpha^2 + 1)^2}{(16\alpha^2 - 1)(16\alpha^2+3)(16\alpha^2 + 5)}$\\
     \midrule
     $(3,0)$  & $-\dfrac{16\alpha^8(4\alpha^2 + 1)^2}{(16\alpha^2 + 1)^2(16\alpha^2 + 3)}$\\
     \bottomrule
    \end{tabular}
    \end{center}
    \caption{All non-zero OPE coefficients up to degree 6 in $d = 4$ for the coulomb gas correlator $\langle V_\alpha V_\alpha V_\alpha V_\alpha \rangle$ in the $s$-channel with $\alpha \equiv Q/4$ to satsify charge neutrality.}
    \label{tab:OPE4d}
\end{table}
\end{comment}

\subsection{Details of analysis of the crossing equation for Abelian $G$-functions}

To explore constraints on the Abelian $G$ function~\eqref{eq:GfunctionAssumed} imposed by crossing symmetry, we substitute this function into the crossing equation Eq.~\eqref{eq:stCrossing}. This gives
\begin{align}
& u^{\frac{\Delta^{(s)}}{2} - \frac{\Delta_1 + \Delta_2}{2}} \textstyle\sum_{n,m \ge 0} C_{nm} v^{\sigma^{(s)}_m} u^{n}
%\nonumber \\ & 
= v^{\frac{\Delta^{(t)}}{2} - \frac{\Delta_2 + \Delta_3}{2}} \textstyle\sum_{n,m \ge 0} C_{nm} u^{\sigma^{(t)}_m} v^{n}.
\label{eq:s-t-crossing}
\end{align}
This equation enforces a structure on the $G$-function that is 
can be described in terms of crossing symmetric building blocks~\cite{Li-InvBootstrap-2017}. Namely, if we require that any truncation up to $(N, M)$ of the double sum in Eq.~\eqref{eq:GfunctionAssumed} is also crossing symmetric,
then all terms in Eq.~\eqref{eq:GfunctionAssumed} fall into two groups: single terms that are crossing-symmetric by themselves, and pairs of terms that get exchanged under crossing. The result is the $G$-function~\eqref{eq:ChannelDependentG} from the main text.

\subsection{Details of analysis of Eq.~\eqref{eq:DeltaConstraint}}

After the identifications $\Delta^{(s)} \equiv \Delta_{\gamma_1 + \gamma_2}$, $\Delta^{(t)} \equiv \Delta_{\gamma_3 + \gamma_2}$, and $\Delta_4 = \Delta_{-\rho_b - \gamma_1 - \gamma_2 - \gamma_3} = \Delta_{\gamma_1 + \gamma_2 + \gamma_3}$, the constraint~\eqref{eq:DeltaConstraint} becomes 
\begin{align}
\label{}
\Delta_{\gamma_1 + \gamma_2} + \Delta_{\gamma_3} - \Delta_{\gamma_1 + \gamma_2 + \gamma_3} + 4\mathcal{Q} 
= 2(\Delta_{\gamma_1} + \Delta_{\gamma_3} - \Delta_{\gamma_3 + \gamma_2})(1 + \mathcal{P}).
\end{align}
Now the continuity of MF spectra allows us to choose $\gamma_1 = \gamma_2 = \epsilon e_i$, where $e_i = (0,\ldots, 1, \ldots, 0)$ (unit in the $i$-th place), with $\epsilon \ll 1$, and $\gamma_3 = \gamma$ in the above equation, which gives
\begin{align}
\label{eq:DeltaConstraint-eps}
&\Delta_{2\epsilon e_i} + \Delta_\gamma - \Delta_{\gamma + 2\epsilon e_i} + 4\mathcal{Q} 
%\nonumber \\ & \quad 
= 2(\Delta_{\epsilon e_i} + \Delta_\gamma - \Delta_{\gamma + \epsilon e_i})(1 + \mathcal{P}).
\end{align}
At this point we expand in powers of $\epsilon$. At order $\epsilon^0$ we get $4\mathcal{Q} = (1 + 2\mathcal{P})\Delta_{0} = 0$, because $\Delta_0 = 0$ by definition. At order $\epsilon$ we get $\mathcal{P} \partial_{q_i}(\Delta_0 - \Delta_\gamma) = 0$ which implies either $\mathcal{P} = 0$ or $\partial_{q_i} \Delta_0 = \partial_{q_i} \Delta_\gamma$. The second condition [together with Eq.~\eqref{Weyl-symmetry}] yields the trivial spectrum $\Delta_\gamma = \text{const} = 0$. Hence, we consider the case $\mathcal{P} = 0$. Then at order $\epsilon^2$ we get $\partial_{q_i}^2(\Delta_0 - \Delta_\gamma) = 0$, which implies that $\Delta_\gamma$ is a quadratic polynomial in all $q_i$. Any such polynomial that vanishes at $\gamma = 0$ and satisfies the symmetry properties~\eqref{Weyl-symmetry} is exactly of the form given in Eq.~\eqref{parabolic-spectrum}~\cite{Karcher2021}. 

\end{document}